\documentstyle[multicol,prl,aps,epsf]{revtex}

\begin{document}
\draft
\tighten

\title{Spin Transport in a Quantum Wire}
\author{Cristina Bena, Leon Balents}

\address{Department of Physics
 University of California Santa Barbara CA 93117}
\date{\today}
\maketitle

\begin{abstract}
We study the effect of electron-electron backscattering interactions
on spin transport in a quantum wire. Even if these interactions have no 
significant effect on charge transport,
they strongly influence the transport of spin. 
We use 
the quantum Boltzmann equation in the collision approximation 
to derive equations of motion for 
spin current and magnetization. In the limit of small
perturbations from equilibrium, we explain 
the existence of `precessional' and `diffusive' behaviors.
We also discuss the low-temperature 
{\it non-linear} decay of an uniform spin current
outside the hydrodynamic regime.
\end{abstract}
\pacs{PACS: 71.10.Pm, 72.10.-d, 75.70.Pa}
\begin{multicols}{2}
\section{Introduction}
Recent developments in the study of one-dimensional (1D) physics 
have been aimed at observing the somewhat intriguing processes of
spin-charge separation and charge fractionalization. 
The theoretical study of various interacting 1D systems 
predicts the decay of the electron into separate
 spin and charge excitations traveling at different velocities 
(see e.g. \cite{emery}).
Also, the charge of the excitations is expected to 
renormalize in the presence of interactions. This phenomenon is called 
charge fractionalization \cite{fisherglazman,kbf,wen}. 
This strongly non-Fermi liquid behavior is described by 
the Luttinger Liquid theory \cite{luttinger}. 

With the discovery of 
quasi-one-dimensional systems such as quantum Hall effect
edge states, quantum wires,
and more recently carbon nanotubes,
there has been good experimental confirmation of this theory
\cite{glattli,bock}.
However, {\it direct} observation of charge fractionalization or 
spin-charge separation remains a challenge, and it is believed that
spin transport may provide useful insight 
into the detection of the theoretically
predicted separation between charge and spin 
\cite{balentsegger1,balentsegger2,spincharge}.
For example, as seen here, 
backscattering interactions terms markedly affect spin transport, 
but do not affect charge transport.

Besides providing insight into the physics of Luttinger
Liquids, spin transport has also  useful technological 
applications \cite{spinexp}. Because of the long 
lifetime of spin excitations,
the spin of the electron may be useful for information
storage or as a transport element, and could have significant 
applications in fields such as quantum computation \cite{quantumcomp}.

This paper investigates the effect of electron-electron
interactions  on spin transport in 1D. 
We consider the model  
of an ideal, infinitely long quantum wire with 
two transport channels: spin up and spin down.
As is well-known, the forward scattering interactions are
 responsible for charge fractionalization and spin-charge
separation, and significantly affect charge transport. By contrast, the
backward scattering processes are marginally irrelevant under the
renormalization group flow \cite{gogolin}, and do not
contribute to the transport of charge. However, the effect of 
backscattering processes on the spin dynamics  
can be quite dramatic. It yields two 
different regimes \cite{balentsegger1}: one characteristic for small
temperatures where the magnetization
vector spatially precesses about the direction of the spin current, and 
another one relevant for higher temperatures where spin diffusion
dominates.
In our analysis we will only take into
account the effects of the backscattering processes; due to spin-charge
separation, we expect these results to be unchanged by the forward
scattering.

Equations of motion for the spin current and magnetization
have been derived in \cite{balentsegger1},
mostly phenomenologically. Here we provide a more general approach.
The treatment we use follows closely
the Kadanoff-Baym formalism \cite{kadanoff}, generalized to
include the electron spin and chirality. We use the quantum Boltzmann
equation in the collision approximation to write down a general form for 
the spin transport equations.

In the {\it hydrodynamic} limit of small perturbations from equilibrium, 
we retrieve the behavior predicted in \cite{balentsegger1}.
In this limit we can identify
the terms responsible for the precession of the magnetization and for 
the spin diffusion. 
We find the spin diffusion coefficient 
\begin{eqnarray}
D_s={ {2 \pi  \hbar^3 v_F^4} \over {u^2  k_B T }},
\nonumber
\end{eqnarray}
in qualitative agreement with 
the estimate of 
Ref.\cite{balentsegger1}, but with a different numerical prefactor. 
This linearized hydrodynamics is manifestly spin-charge separated, giving
us some confidence in the validity of the result.

In the presence of larger perturbations,
non-linear terms appear as well.
An analysis of the full equations of motion 
may be done numerically. For the particular case of an injected
uniform spin current, we perform a numerical analysis
of the equations at various temperatures. At high 
temperatures, we retrieve purely diffusive behavior. 
In the limit of zero temperature, however, the system
ceases to show exponentially decaying behavior and 
we can find an analytic solution for the decay of the spin current.
This solution is in very good agreement with the corresponding
numerical results.

We check that the backscattering interactions cause
no corrections to the charge and heat transport equations.

A comment on the validity of these results is in order. 
It seems quite likely that
the quantum Boltzmann approach is asymptotically exact in the
hydrodynamic regime, due to the marginal irrelevance of backscattering in
equilibrium (up to possible logarithmic corrections requiring a higher-order
analysis). The low-temperature limit is, however, much less transparent.
Indeed, the quantum Boltzmann 
description is almost certainly not exact in this regime.
In particular, while the original formulation is manifestly spin-charge
separated, the quantum Boltzmann equations are not, 
but become so in linear response. Thus the decay of the spin
current found at low temperatures depends upon the form of
the charge distribution functions defined in Section IV. Nevertheless,
this dependence appears fairly innocuous, and leads to no obvious 
unphysical behavior for local quantities (e.g. magnetization,
charge density, current, energy density). For this reason, and since the
results are in agreement with expectations based on scaling,
we find the quantum Boltzmann approximation reasonable an this limit as well.

In the following section 
we write down the 
spin transport equations. The derivation of these equations is presented 
in detail in the Appendix. In section III, 
we study the perturbative regime and the resulting
precessional and diffusive behaviors. In section IV, using numerical 
and analytical methods, we study the time
evolution of a uniform spin current injected into the system.
We present the conclusions in section V.

\section{Spin transport equations}

The non-interacting low energy effective Hamiltonian for a
spin-$1 \over 2$ quantum wire is given by:

\begin{equation}
H_0=\sum_{\alpha}\int\! dx\ \hbar v_F [\psi^{\dagger}_{L\alpha}i\partial_x
\psi^{\vphantom \dagger}_{L\alpha}-
\psi^{\dagger}_{R\alpha}i\partial_x\psi^{\vphantom \dagger}_{R\alpha}],
\label{h0}
\end{equation}
where $\alpha=\uparrow / \downarrow$ is the spin index, $R/L$ 
denotes the chirality of the right- and left- moving modes, and $v_F$ 
is the Fermi velocity. The quantum wire is taken to be infinitely long.
We consider adding two types of local interactions that
are allowed by $SU(2)$ invariance. The forward scattering
interaction is described by:

\begin{eqnarray}
H_1&&=
-u_f \int dx J_R(x) J_L(x)
\label{h1} \\
&&=-u_f \sum_{\alpha,\beta} 
\int dx \psi^{\dagger}_{R\alpha}(x) \psi_{R\alpha}(x)
\psi^{\dagger}_{L\beta}(x) \psi_{L\beta}(x),
\nonumber
\end{eqnarray}
while the backward scattering interaction is:
\begin{eqnarray}
&&
H_2=-u \int dx \vec{\bf J}_R(x) \cdot \vec{\bf J}_L(x)
\label{h2} \\ &&=-{u \over 4} \sum_{\alpha,\beta, \gamma, \delta} 
\int dx \psi^{\dagger}_{R\alpha}(x) 
\vec{\sigma}_{\alpha\beta}\psi_{R\beta}(x) \cdot 
\psi^{\dagger}_{L\gamma}(x) \vec{\sigma}_{\gamma\delta} \psi_{L\delta}(x).
\nonumber
\end{eqnarray}
Of the quartic processes described by $H_1$ and $H_2$, only the 
backscattering processes will contribute to 
spin transport. The forward scattering terms are included as a check.

Given the above form for the interactions, 
we are interested in deriving equations for
the generalized Wigner distribution functions
$f^{\alpha_1 \alpha_2}_{r}(p;x,t)$.
For each chirality, these functions form a $2 \times 2$  
matrix in the spin space.
If we choose the $z$ axis as the reference axis,
the diagonal components of this matrix give the distribution 
functions for the particles with the spin directed upward and 
downward along this axis. 
The off-diagonal components are related to the distributions of particles
that have the spin directed along the $x$ and $y$ axes.
In equilibrium the off diagonal components vanish, while
the diagonal ones are equal and independent of space and time, 
and the Wigner matrix can be written as:

\begin{eqnarray}
f_r^{\alpha \beta}(p;x,t)=f^0_r(p) \delta_{\alpha \beta},
\end{eqnarray}
\begin{eqnarray}
f^0_{R/L}(p)={1 \over {e^{\large \pm {{\hbar p v_F} \over {k_B T}}}+1}}.
\end{eqnarray}

We can relate the Wigner matrices to the actual spin current 
and magnetization,
as well as 
with the total charge density and charge current. The local densities
of the charge and of the spin components 
in the right and left moving channels are given by:

\begin{eqnarray}
{\bf M}^{i}_r({\cal X})={1 \over 2}  \sum_{\alpha_1, \alpha_2} \int_p 
f_r^{\alpha_1 \alpha_2}(p;{\cal X}) \sigma_{\alpha_2 \alpha_1}^{i},
\label{defofj1}
\end{eqnarray}
\begin{eqnarray}
n_r({\cal X})=\sum_{\alpha} \int_p f_r^{\alpha \alpha}(p;{\cal X}),
\label{defofi1}
\end{eqnarray}

and the total spin current and magnetization are

\begin{eqnarray}
{\bf \vec J_s}=v_F({\bf \vec M}_R-{\bf \vec M}_L),
\nonumber
\end{eqnarray}
\begin{eqnarray}
\vec{\bf M}={\vec{\bf M}}_R+{\vec{\bf M}}_L.
\label{defofj2}
\end{eqnarray}
For simplicity
we chose to denote $(x,t)$ by ${\cal X}$ and $\int {dp \over {2 \pi}}$
by $\int_p$, and we will make use of this notation
throughout the paper.

The total charge density and current can be written as
\begin{eqnarray}
I=v_F(n_R-n_L),
\nonumber
\end{eqnarray}
\begin{eqnarray}
\rho=n_R+n_L.
\label{defofi2}
\end{eqnarray}

In equilibrium all the components of the 
spin current and magnetization vanish.
If the system is perturbed, the effect of the interactions is
to move the system towards its equilibrium position.
The equations that describe this dynamics can be 
derived using a technique similar to \cite{kadanoff}. The details of the
derivation are presented in the Appendix.

The resulting generalized Boltzmann equations for the Wigner 
distribution functions $f_r^{\alpha_1 \alpha_2}$
can be written as:
\end{multicols}
\begin{eqnarray}
(\partial_t + v_F \partial_x)f_{R}^{\alpha_1 \alpha_2}(p)
&&= {{u} \over {2 \hbar}} i [f_R^{\alpha_1 \gamma}(p) 
\int_{q} f_L^{\gamma \alpha_2}(q)-\int_{q} 
f_L^{\alpha_1 \gamma}(q) f_R^{\gamma \alpha_2}(p)] 
\label{eqp10}
\\&&+{{u^2} \over {8 \hbar^2 v_F}} 
\bigg\{ 2 f_R^{\alpha_1 \alpha_2}(p)  \bigg[ f_R^{\alpha_4 
\alpha_3}(p) \int_{q} f_L^{\alpha_3 \alpha_4}(q)-\int_q f_L^{\gamma \gamma}
(q) \bigg]
-2\int_q f_L^{\alpha_1 \alpha_2}(q)  \bigg[ f_L^{\alpha_4 \alpha_3}(q) 
 f_R^{\alpha_3 \alpha_4}(p)-f_R^{\gamma \gamma}(p) \bigg]
\nonumber \\&& +\int_q f_L^{\alpha_1 \gamma}(q) f_R^{\gamma \alpha_2}(p)
\bigg[f_L^{\alpha_3 \alpha_3}(q)-f_R^{\alpha_3
 \alpha_3}(p)\bigg]-\int_q f_R^{\alpha_1 \gamma}(p) f_L^{\gamma \alpha_2}(q)
\bigg[f_R^{\alpha_3 \alpha_3}(p)-f_L^{\alpha_3 
\alpha_3}(q)\bigg]\bigg\},
\nonumber
\end{eqnarray}
while the corresponding 
equation for $f_L^{\alpha_1 \alpha_2}$ can be obtained from 
Eq.(\ref{eqp10}) by interchanging $L$ and $R$ everywhere and by replacing 
$(\partial_t+ v_F \partial_x)$ by $(\partial_t- v_F \partial_x)$.
To simplify the notation, 
as all the distribution functions are evaluated at the same position $x$
and at the same time $t$, we dropped the spatial and temporal
coordinates.
\begin{multicols}{2}
\section{Perturbative expansion}

If the deviations from equilibrium are small,
we can write the matrices of Wigner distribution functions as a sum of
equilibrium diagonal matrices and small perturbations,
$f_r^{\alpha \beta}(p;{\cal X})=f^0_r(p) \delta_{\alpha \beta}+ 
F_r^{\alpha \beta}(p;{\cal X})$.
We use an integral form of Eq.(\ref{eqp10}) that is given in the 
Appendix in the equation (\ref{eqforf}). 
We expand this form in $F_r^{\alpha \beta}$,
so that 
in the terms proportional to $u^2$ we keep only the linear terms in 
$F$. We make use of the fact that
$f_R^0(p)+f_L^0(p)=1$ and we obtain
\\
\phantom{1}
\put(-275,-5){\line(1,0){255}}
\put(-20,-5){\line(0,1){10}}
\end{multicols}

\begin{eqnarray}
(\partial_t \pm v_F \partial_x)\int_p f_{R/L}^{\alpha_1 \alpha_2}(p;{\cal X})=
&&\pm {{u} \over {2 \hbar}}
i \sum_{\gamma} \int_{q} F_R^{\alpha_1 \gamma}(q;{\cal X})
\int_p F_L^{\gamma \alpha_2}(p;{\cal X})
\mp i {{u} \over {2 \hbar}} \sum_{\gamma}
\int_q F_L^{\alpha_1 \gamma}(q;{\cal X}) 
\int_p F_R^{\gamma \alpha_2}(p;{\cal X}) 
\label{pertf}
\\&& 
\mp {{u^2} \over {2  \hbar^2 v_F}} \int_{q} 
\Big[ F_R^{\alpha_1 \alpha_2}(q;{\cal X})-
F_L^{\alpha_1 \alpha_2}(q;{\cal X})\Big]
\int_{p} f_R^0(p) f_L^0(p) 
\nonumber \\&&
\pm {{u^2} \over {4  \hbar^2 v_F}} \delta_{\alpha_1 \alpha_2}\sum_{\alpha_3}
\int_{q} \Big[F_R^{\alpha_3 \alpha_3}(q;{\cal X})
-F_L^{\alpha_3 \alpha_3}(q;{\cal X})\Big] \int_p f_R^0(p) f_L^0(p).
\nonumber
\end{eqnarray}
\put(255,0){\line(1,0){255}}
\put(255,-10){\line(0,1){10}}
\begin{multicols}{2}
We use Eq.(\ref{defofj1}) and
\begin{eqnarray}
\int_p f_R^0(p) f_L^0(p)&&=\int {dp \over {2 \pi}}
 {1 \over {e^{{ \hbar p v_F} \over {k_B T}}+1}}
{1 \over {e^{{- \hbar p v_F} \over {k_B T}}+1}}
\nonumber \\&&
={1 \over {2 \pi}} {{k_B T} \over { \hbar v_F}}
\end{eqnarray}

to write
\begin{eqnarray}
(\partial_t \pm v_F \partial_x)&&{\bf M}_{R/L}^{i }({\cal X})= \pm
{{u} \over {2 \hbar}}
\sum_{j,k}{\bf M}_R^{j }({\cal X}) {\bf M}_L^{k}({\cal X}) \epsilon^{j k i}
\nonumber \\&&
\mp {{u^2} \over {4 \pi}} {{k_B T}  \over {\hbar^3 v_F^2}} 
 [{\bf M}_R^{i}({\cal X})-{\bf M}_L^{i}({\cal X})].
\label{res0}
\end{eqnarray}

The temperature dependence of the second term only is a result of  
an explicit dependence of this term 
on the form of the equilibrium distribution functions, while the
first term linear in $u$ is independent of this form.

In terms of the total spin current and magnetization Eq.(\ref{res0}) becomes: 
\begin{eqnarray}
\partial_t \vec{\bf J}_s({\cal X}) +v_F^2 
\partial_x \vec{\bf M}&&({\cal X})=
{u \over \hbar} {\bf \vec{M}} ({\cal X}) \times 
\vec{\bf J}_s ({\cal X})-
\nonumber \\&&-u^2 {{ k_B T} \over { 2 \pi  \hbar^3 v_F^2}}
\vec{\bf J}_s({\cal X}).
\label{result}
\end{eqnarray}

We also note that the spin continuity equation still holds:
\begin{equation}
\partial_t \vec{\bf M}({\cal X})+\partial_x \vec{\bf J}_s({\cal X})=0.
\end{equation}

Equation (\ref{result}) is qualitatively similar to the corresponding result 
in \cite{balentsegger1}, up to a numerical 
factor in front of the $\vec{\bf J}_s$ term.      

The charge dynamics is unaffected. We obtain from 
Eqs.(\ref{defofi1},\ref{defofi2},\ref{pertf}):
\begin{eqnarray}
\partial_t \rho({\cal X})+\partial_x I({\cal X})=0,
\end{eqnarray}
\begin{eqnarray}
\partial_t I({\cal X})+v_F^2 \partial_x \rho({\cal X})=0.
\end{eqnarray}
As expected, this result holds as well 
in the non-perturbative regime, and it can be exactly derived using
the more general equation (\ref{eqforf}).

We analyze Eq.(\ref{result}) 
to extract the physics of the system. If the first term of the
RHS of the equation dominates, 
as it is the case for zero or small temperatures, the spin transport is
ballistic. In the low frequency linear response limit, the
time derivative of $\vec{\bf J}_s$ can be neglected. The spin current is
thus constant in time, and the space dependence of the 
magnetization is a precession about the direction of the 
spin current \cite{balentsegger1,balentsegger2}.

In the opposite limit of large temperature, the second term will
dominate and the spin transport will be mainly diffusive.  
We retrieve the diffusion equation:

\begin{equation}
v_F^2 \partial_x^2 \vec{\bf M}({\cal X})+
u^2 {{ k_B T} \over { 2 \pi  \hbar^3 v_F^2}}
\partial_t \vec{\bf M}({\cal X})=0,
\label{pert}
\end{equation}

and we thus extract the spin diffusion coefficient:

\begin{equation}
D_s={ {2 \pi  \hbar^3 v_F^4} \over {u^2  k_B T }}.
\end{equation}
This differs only by a factor of $8$ 
from the value derived in \cite{balentsegger1}.

The crossover temperature between
diffusive and precessional spin transport can be estimated
as a function of the interaction strength and of magnetization

\begin{equation}
k_B T_{crossover} \approx {\bf M} {2 \pi \hbar^2 v_F^2 \over {u}}.
\end{equation}

\section{Time evolution of an uniform spin current}

Here we find the time evolution
of a spatially uniform 
spin current with the only non-zero component on the $z$ axis,
injected into the system at $t=0$. We do not use a perturbative
approach as in the previous section, but we focus on the
more general form in Eq.(\ref{eqp10}) and we try to study it both 
numerically and analytically.
We assume the initial magnetization is zero everywhere. 
This can be implemented by
applying different chemical potentials on the right/left,
spin up/down subspecies of particles. In particular we will choose
chemical potentials $\mu$ for the spin up right movers
and for the spin down left movers and respectively chemical potentials 
$-\mu$ for the spin down right movers
and for the spin up left movers. The initial Fermi functions for the 
electrons are thus space independent and can be written as:

\begin{eqnarray}
f_R^{\uparrow \uparrow}(p,t=0,x)=
{1 \over {e^{{{\hbar p v_F} \over {k_B T}}-\mu}+1}},
\label{in0}
\\
f_R^{\downarrow \downarrow}(p,t=0,x)=
{1 \over {e^{{{\hbar p v_F} \over {k_B T}}+\mu}+1}},
\label{in1}
\\
f_L^{\uparrow \uparrow}(p,t=0,x)=
{1 \over {e^{-{{\hbar p v_F} \over {k_B T}}+\mu}+1}},
\label{in2}
\\
f_L^{\downarrow \downarrow}(p,t=0,x)=
{1 \over {e^{-{{\hbar p v_F} \over {k_B T}}-\mu}+1}},
\label{in3}
\\
f_R^{\alpha_1 \alpha_2}(p,t=0,x)=0 , {\text{ for all  }} \alpha_1 \ne \alpha_2.
\label{in4}
\end{eqnarray}

For this particular case, we can write simplified equations involving only the
diagonal components of the distribution function matrix.
From Eq.(\ref{eqp10}) we derive
\\
\phantom{1}
\put(-275,-10){\line(1,0){255}}
\put(-20,-10){\line(0,1){10}}
\end{multicols}
\begin{eqnarray}
\partial_t f_{R}^{{\uparrow \uparrow}/{\downarrow \downarrow}}(p)=
\pm {{u^2} \over {4 \hbar^2 v_F}}
\Big\{ &&f_{R}^{\uparrow \uparrow}(p) f_{R}^{\downarrow \downarrow}(p)
\int_q [f_{L}^{\downarrow \downarrow}(q)-f_{L}^{\uparrow \uparrow}(q)]+
\int_q f_{L}^{\uparrow \uparrow}(q) f_{L}^{\downarrow \downarrow}(q)
[f_R^{\uparrow \uparrow}(p)-f_{R}^{\downarrow \downarrow}(p)]+
\nonumber \\&&
+\int_q f_{L}^{\uparrow \uparrow}(q) f_{R}^{\downarrow \downarrow}(p)-
 f_{R}^{\uparrow \uparrow}(p) \int_q f_{L}
^{\downarrow \downarrow}(q) \Big\},
\label{eqR}
\end{eqnarray}
\put(255,0){\line(1,0){255}}
\put(255,-10){\line(0,1){10}}

\begin{multicols}{2}
The equations for the distributions functions 
of the left movers are obtained simply by interchanging
$L$ and $R$ everywhere in the above formula.
These equations can be solved numerically for finite temperatures
when we impose the initial conditions from Eqs.(\ref{in0} - \ref{in4}).

\subsection{High temperature regime}
In the limit of temperatures that are much larger than the applied
chemical potential, the numerical analysis indicates that the system evolves
towards an equilibrium state described by the usual Fermi
distributions.

We can compare the numerical results with an analytical estimate by
noting that in this limit the quantum Boltzmann equations for the
system are easily solvable. This is because
at each momentum, the variations from the
equilibrium distribution function are small, and we can describe
the system by the equations derived in section III, specialized to
the case of zero magnetization. From Eq.(\ref{pert})
we derive: 
\begin{equation}
\partial_t {\bf J}_s^z=-u^2 {{k_B T}\over{2 \pi \hbar^3 v_F^2}} {\bf J}_s^z.
\end{equation}
with the solution
\begin{equation}
{\bf J}_s^z(t)={\bf J}_s^z(0) e^{-t {\cal D}},
\end{equation}
where ${\cal D}={{ u^2 k_B T}\over {2 \pi \hbar^3 v_F^2}}$.

This is confirmed by the numerical analysis.
When we plot the logarithm of the current as a function of time
we obtain straight lines, consistent with the assumption that
the current is decaying exponentially. If we set all the constants
to $1$ the slope of the decay is just $-u^2 T$. 
We compare the numerical results with the 
theoretical estimates for an exponential decay (represented in the figure
by the dashed lines) and the results are in very good agreement.
\narrowtext
\begin{figure}
\epsfxsize=2.6in
\epsfysize=2.9in
\centerline{\epsffile{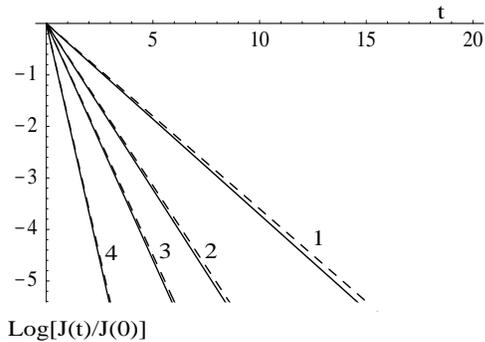}}
\caption{The logarithm of the spin current plotted as a function of time.
The interaction strength has been taken equal to $0.3$, the chemical potential
$\mu=1$ and the curves denoted by $1$, $2$, $3$ and $4$ correspond to the
temperatures of $T=4$, $7$, $10$ and $20$ respectively.}
\label{fig1}
\end{figure}

\subsection{Zero temperature regime}

\begin{figure}
\epsfxsize=2.8in
\epsfysize=3.4in
\centerline{\epsffile{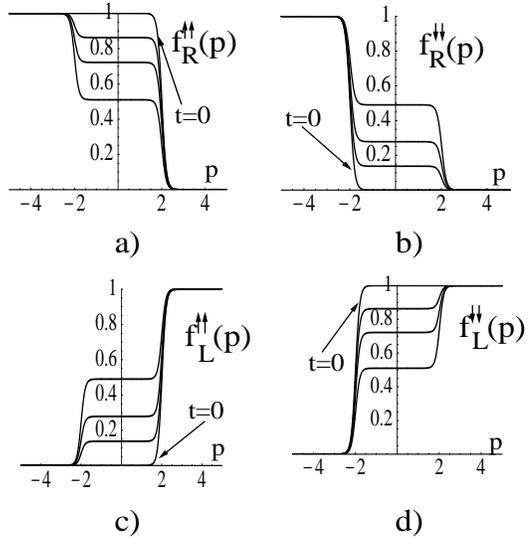}}
\caption{The distribution functions for a) spin up right movers,
b) spin down right movers c) spin up left movers d) spin down left movers
depicted at $t=0$ and at subsequent times. After sufficiently long times,
the values for all the individual distribution functions will be constant
in the region $-\mu/\hbar v_F < p< \mu/\hbar v_F$ 
and equal to $1/2$. Here we set all the
constants to $1$ and we chose 
$\mu =2$, $T=0.1$ and $u=0.3$}
\label{fig2}
\end{figure}

In the opposite limit of zero temperature (or $\mu \gg k_B T$) the numerical
analysis indicates strong deviations from the purely exponential decay.
This can be understood as follows: at low temperatures 
the perturbations cannot be treated 
as small with respect to the equilibrium Fermi function. As an example,
in Fig.\ref{fig2} we plot the distribution functions for the
case we described in Eqs.(\ref{in0}-\ref{in4}), at $t=0$ and our 
numerical predictions for the form of these distribution functions at 
later times. In the range $|p| < \mu/\hbar v_F$ the
difference from the equilibrium distributions 
can be as high as $1$, therefore making
the perturbative approach of section III inconsistent.

We note that instead of moving toward
the unperturbed Fermi distributions, in the region $ |p| < \mu/\hbar v_F$
the value of the distribution functions is a constant. At very large times
this constant goes to $1/2$.
This type of behavior is entirely consistent with the physics of the
system. In one dimension with linear dispersion, for the specific form of
the interactions we consider, two interacting electrons
can only transfer spin while the momenta of the electrons are
preserved. Thus, the sum of the numbers of spin up and spin down right 
moving electrons at each momentum has to remain constant in time. 

Taking this observation into account, we can also solve analytically
the equations of motion for the spin current in the limit of zero temperature.
We define $f_R^s={1 \over 2}
[f_R^{\uparrow \uparrow}-f_R^{\downarrow \downarrow}]$
and $f_R^c=f_R^{\uparrow \uparrow}+f_R^{\downarrow \downarrow}$. 
In these new variables Eq.(\ref{eqR}) becomes:

\begin{eqnarray}
\partial_t f_{R/L}^s&&(p)={{u^2} \over {4 \hbar^2 v_F}} 
\Big\{\int_q f_{L/R}^s(q)\Big[f_{R/L}^c(p)-{1 \over 2} [{f_{R/L}^{c}}(p)]^2
\Big]
\label{eqS1}
\\&&-f_{R/L}^s(p)\Big[\int_q f_L^c(q)-{1 \over 2} 
\int_q [{f_{L/R}^{c}}(p)]^2\Big]
\nonumber \\&&+2 [f_{R/L}^{s}(p)]^2 
\int_q f_{L/R}^s(q)-2 \int_q [f_{L/R}^{s}(q)]^2 f_{R/L}^s(p)\Big\},
\nonumber
\end{eqnarray}

\begin{figure}
\epsfxsize=3in
\epsfysize=3.6in
\centerline{\epsffile{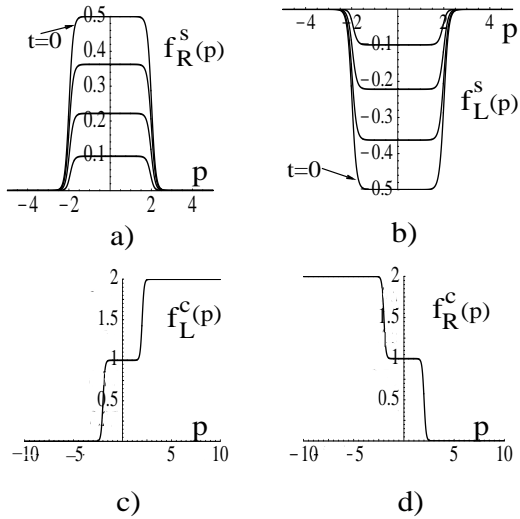}}
\caption{The distribution functions a)$f_R^s$, b)$f_L^s$, c)$f_L^c$, 
d)$f_R^c$ at $t=0$ and at subsequent times. The value for $\mu$ has 
been set to $2$ while the value for $u$ is $0.3$. We note that the
``charge'' distribution functions are time independent, while the ``spin''
distributions functions are constant on the interval $(-\mu, \mu)$, and 
decrease to zero after long enough times.}
\label{fig3}
\end{figure}

For the case defined by Eqs.(\ref{in0} -\ref{in4}),
the form of the analytical solution can be inferred from the 
results of the numerical analysis.
In Fig.\ref{fig3} we plot $f_{R/L}^{c/s}$ at $t=0$ and subsequent
times as obtained from the numerical analysis.

Using Eq.(\ref{eqR}), we note that $\partial_t f_{R/L}^c=0$. 
Therefore these distribution functions are constant in time and at any time
we can use for them the values we fixed at $t=0$. 
For this particular case,
\begin{eqnarray}
f_R^c(p)=\Theta(-p+{\mu \over {\hbar v_F}})+
\Theta(-p-{\mu \over {\hbar v_F}})
\end{eqnarray} 
and
\begin{eqnarray}
f_L^c(p)=\Theta(p-{\mu \over {\hbar v_F}})+
\Theta(p+{\mu \over {\hbar v_F}}).
\end{eqnarray} 
Then Eqs.(\ref{eqS1}) simplify, so that for $|p|<\mu/\hbar v_F$:
\begin{eqnarray}
\partial_t f&&_{R/L}^s(p)={{u^2} \over {4 \hbar^2 v_F}} 
\Big\{
{1 \over 2} \int_q f_{L/R}^s(q)- {\mu \over {2 \pi \hbar v_F}} f_{R/L}^s(p)
\label{eqs3}\\&&+2 [f_{R/L}^{s}(p)]^2 
\int_q f_{L/R}^s(q)-2 \int_q [f_{L/R}^{s}(q)]^2 f_{R/L}^s(p)\Big\},
\nonumber
\end{eqnarray}
while
\begin{equation}
\partial_t f_{R/L}^s(p) =0
\label{eqs31}
\end{equation}
for other values of $p$.

From Eqs.(\ref{eqs3},\ref{eqs31}) 
as well as from the numerical analysis plotted
in Fig.\ref{fig3}, 
we note that at $t=0$, $f_R^s(p,t)+f_L^s(p,t)=0$, which implies
that $f_R^s(p,t)+f_L^s(p,t)=0$ at all times. 
Moreover we can pick a solution of the form 
\begin{eqnarray}
f_{R/L}^s(p,t)=\bigg\{ \begin{array}{c}
\pm \lambda(t), {\text{for }}|p|<\mu/\hbar v_F, \\ 0, 
{\text{    otherwise.}} \end{array} 
\nonumber
\end{eqnarray}
with $\lambda(0)=1/2$. Therefore the equation for $\lambda$ becomes:
\begin{equation}
{{d \lambda} \over {d t}}=-{{u^2 \mu} \over {4 \pi \hbar^3 v_F^2}}(\lambda 
+4 \lambda^3.)
\end{equation}
Noting that ${\bf J}_z=\int_p[f_R^s(p,t)-f_L^s(p,t)]=4 \lambda(t) \mu/
2 \pi \hbar v_F$,
we obtain a time dependence for the spin current of the form:

\begin{equation}
{\bf J}^z(t)={{{\bf J}^z(0)} \over {\sqrt{2 e^{t/t_0}-1}}},
\label{t0}
\end{equation}
where ${\bf J}^z(0)=2 \mu/ 2 \pi \hbar v_F$ 
and $t_0={2 \pi \hbar^3 v_F^2}/{\mu u^2}$.

This result agrees very well with the numerical analysis 
for all values of the parameters in the system. 
Below we plot the numerical versus the analytical result, for 
a particular set of values for $T$, $\mu$ and $u$.

\begin{figure}
\epsfxsize=2.5in
\epsfysize=2.4in
\centerline{\epsffile{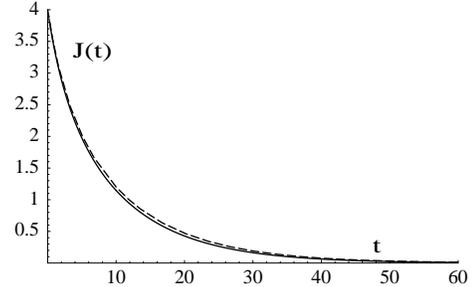}}
\caption{The spin current is plotted as a function of time, for
$T=0.1$, $\mu=2$ and $u=0.3$. Both the 
numerical and analytical result are given; the dashed lines corresponds
to the estimate of Eq.(\ref{t0}), while the full line is the numerical
result.}
\label{fig4}
\end{figure} 

We note that for short times ($t \ll t_0$)
the form of the decay is a power law
$ {\bf J}^z(t)={{{\bf J}^z(0)} \over {\sqrt{1+2{t/t_0}}}}$, while for large 
enough times $t \gg t_0$ we retrieve a purely exponential decay 
$ {\bf J}^z(t)={{{\bf J}^z(0)} \over \sqrt{2}} e^{-t/2 t_0}$

\subsection{Intermediate temperature regime}

We can perform the numerical analysis for the intermediate regime of
temperature when the temperature is comparable to the applied 
chemical potential.
We study the crossover between the 
high temperature and low temperature regime by plotting the 
rate of change ${\cal R}$ 
of the log of the current at large enough time scales
as a function of $\mu$ or $T$. 
As seen in the previous sections, 
in the zero temperature limit we expect this to be a constant 
${\cal R}=-1/2 t_0 \sim -\mu u^2/2$, 
while in the high temperature limit ${\cal R} =
-{\cal D} \sim -u^2 T$

\begin{figure}
\epsfxsize=3.5in
\epsfysize=4.2in
\centerline{\epsffile{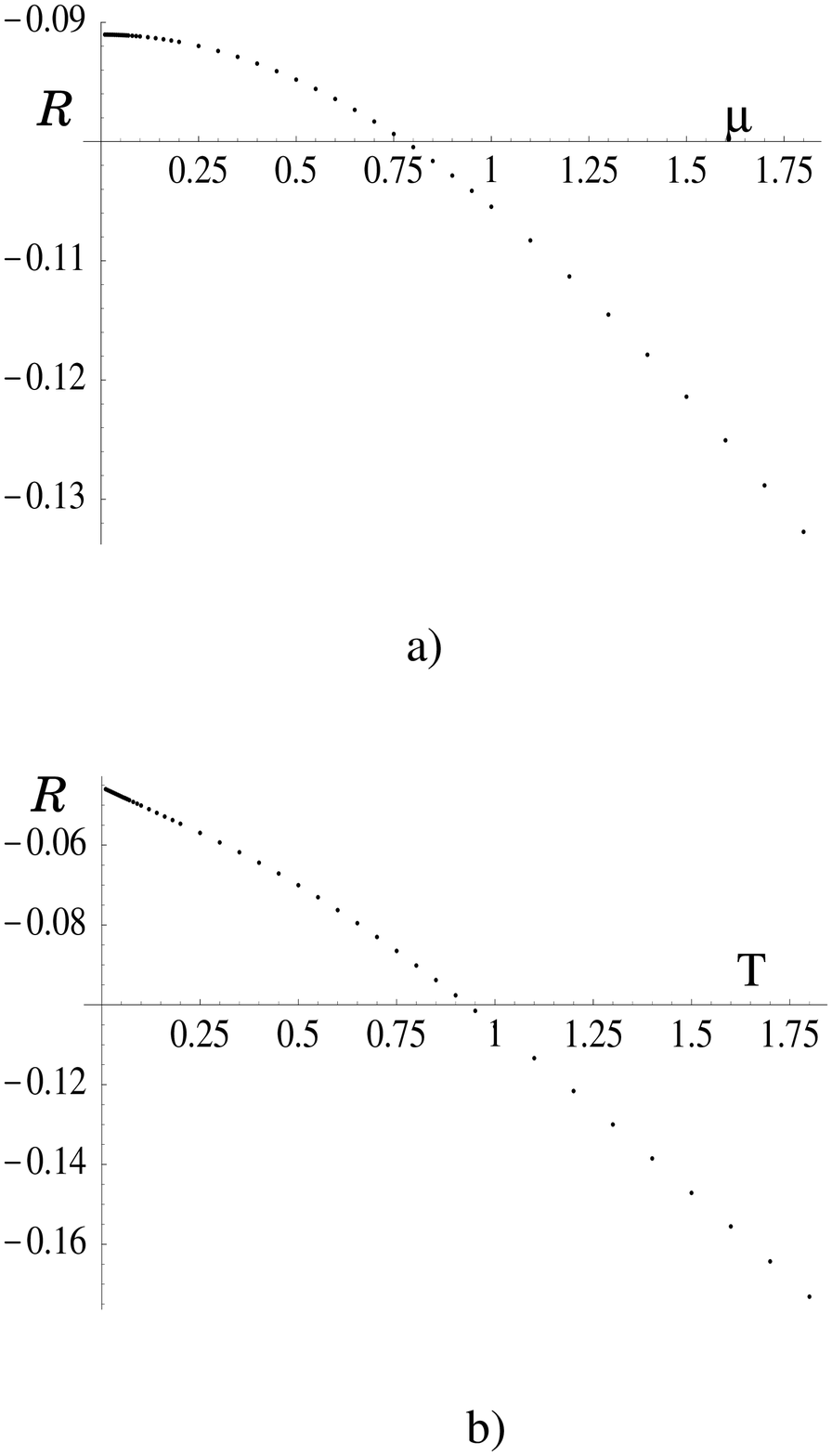}}
\caption{The rate of change of the log of the spin current with respect to
\\
a) the applied chemical potential $\mu$ when the temperature is held
constant $T=1$.
\\
b) the temperature $T$, when the chemical potential is held constant to
$\mu=1$.\\
The value for $u$ is $0.3$, $1/t_0=0.09$, and ${\cal D}=0.09$.}
\label{fig5}
\end{figure}

In Fig.\ref{fig5} a) we plot 
 ${\cal R}$ as a function of $T$, while keeping the chemical 
potential fixed to $\mu=1$, and in b)
we plot ${\cal R}$ as a function of
of $\mu$ while keeping the temperature constant $T=1$.

As we showed before, when $T$ is held fixed
and $\mu$ is small
$R$ goes to a constant value ${\cal D} \sim -u^2 T =-0.09$. Similarly,
when $\mu$ is fixed
and $T$ is small
$R$ becomes  -$1/2 t_0 \sim -u^2 \mu/2 =-0.045$.
The figures above also portray the crossover between the two regimes.

\section{Conclusion}

We analyzed the effects of backscattering interactions on 
spin transport in a one dimensional quantum wire. 
Using the quantum Boltzmann equation we wrote down the equations of
motion for the spin current and magnetization.
We found that in the hydrodynamic limit of small perturbations, the 
backscattering processes generate terms in the equation
of motion responsible for bulk precession of magnetization as well as for
spin diffusion. We computed the diffusion coefficient.
 
We also analyzed both numerically and analytically 
the time evolution of a uniform spin current injected into the 
system. In the limit of high temperatures, this shows
an exponential decay controlled by the values of the
temperature and of the interaction strength. 
In the zero temperature limit, the analytic 
results indicate a more complex behavior characterized by a
transition between a power law behavior
at small times to an exponential decay behavior controlled by the 
strength of the applied chemical potential and by the strength of
the interactions, at large enough times.

\section{Acknowledgements}
This research has been supported by NSF-DMR-9985255, the Sloan and 
Packard foundations, and by Ferrando-Fithian Fellowship and 
Parsons Foundation Fellowship.

\end{multicols}
\widetext

\appendix
\section{}
\subsection{Equations of motion for the imaginary time Green's function}

Given the non-interacting low energy effective Hamiltonian for a
spin-$1 \over 2$ quantum wire described in Eq.(\ref{h0}):

\begin{equation}
H_0=\sum_{\alpha}\int\! dx\ \hbar v_F [\psi^{\dagger}_{L\alpha}i\partial_x
\psi^{\vphantom \dagger}_{L\alpha}-
\psi^{\dagger}_{R\alpha}i\partial_x\psi^{\vphantom \dagger}_{R\alpha}],
\end{equation}

and the local interactions defined in Eqs.(\ref{h1},\ref{h2}):

\begin{eqnarray}
H_1=
-u_f \int dx J_R(x) J_L(x)
=-u_f \sum_{\alpha,\beta} 
\int dx \psi^{\dagger}_{R\alpha}(x) \psi_{R\alpha}(x)
\psi^{\dagger}_{L\beta}(x) \psi_{L\beta}(x),
\end{eqnarray}
and
\begin{eqnarray}
H_2=-u \int dx \vec{\bf J}_R(x) \cdot \vec{\bf J}_L(x)
=-{u \over 4} \sum_{\alpha,\beta, \gamma, \delta} 
\int dx \psi^{\dagger}_{R\alpha}(x) 
\vec{\sigma}_{\alpha\beta}\psi_{R\beta}(x) \cdot 
\psi^{\dagger}_{L\gamma}(x) \vec{\sigma}_{\gamma\delta} \psi_{L\delta}(x),
\end{eqnarray}
 
we derive the transport equations for the spin.

We start with
the equations for $\psi_{r \alpha}$:

\begin{eqnarray}
i \hbar \partial_t \psi_{r \alpha}(x,t)=[H(t),\psi_{r \alpha}(x,t)],
\end{eqnarray}

which implies

\begin{eqnarray}
i \hbar [\partial_t -\hbar v_F (-1)^{\epsilon_r}
\partial_x ]\psi_{r \alpha}(x,t)=&&
\Big[-{{u} \over 2} \delta_{r R} 
\sum_{\gamma} \psi^{\dagger}_{L \gamma}(x,t) \psi_{L \alpha}(x,t)
\psi_{R \gamma}(x,t)
\\&&
+({{u} \over 4}-u_f) \delta_{r R} 
\sum_{\gamma} \psi^{\dagger}_{L \gamma}(x,t) \psi_{L \gamma}(x,t)
\psi_{R \alpha}(x,t)\Big]+(R \leftrightarrow L),
\label{eqforpsi}
\nonumber
\end{eqnarray}

where $r=R/L$,
$\epsilon_R=1$ and $\epsilon_L=2$. 

We define the imaginary time Green's functions to be: 

\begin{eqnarray}
G^{\alpha_1 \alpha_2}_{r_1 r_2} ({\cal X}_1;{\cal X}_2)=
-i {{< T [S \psi_{r_1 \alpha_1}({\cal X}_1)
{\psi_{r_2 \alpha_2}^{\dagger}}({\cal X}_2)]>}
\over {<T[S]>}},
\end{eqnarray}

\begin{eqnarray}
 G_2({{\cal X}_1}_{r_1}^{\alpha_1},{{\cal X}_2}_{r_2}^{\alpha_2},
{{\cal X}_3}_{r_3}^{\alpha_3},{{\cal X}_4}_{r_4}^{\alpha_4})
=-{{<T[S \psi_{r_1 \alpha_1}({\cal X}_1)
\psi_{r_2 \alpha_2}({\cal X}_2) \psi^{\dagger}_{r_4 \alpha_4}({\cal X}_4)
\psi^{\dagger}_{r_3 \alpha_3}({\cal X}_3)]>} \over {<T[S]>}},
\nonumber
\end{eqnarray}

where $\alpha_1, \alpha_2 =
\uparrow/\downarrow$, $r_{i}=R/L$, and ${\cal X}_i$
stands for $x_i,t_i$. Also,
$T$ denotes imaginary time ordering, and
$S$ is given by

\begin{eqnarray}
S=\exp{\Big[-{i\over{\hbar}} \sum_{\sigma, \gamma=\uparrow/\downarrow}}
&& {\sum_{r,r'} \int dx \int_0^{-i \beta} dt 
U_{r r'}^{\sigma \gamma}({\cal X})
 \psi^\dagger_{\sigma  r}({\cal X})
\psi_{\gamma r'}({\cal X})\Big]}. 
\nonumber 
\end{eqnarray}

Using Eq.(\ref{eqforpsi}) we can derive the equations of motions for 
the imaginary time Green's functions:

\begin{eqnarray}
 \hbar[ \partial_{t_1}-&&(-1)^{\epsilon_{r_1}} v_F \partial_{x_1}]
G_{r_1 r_2}^{\alpha_1 \alpha_2}({\cal X}_1,{\cal X}_2)+
i \sum_{r,\alpha} U_{r_1 r}^{\alpha_1 \alpha}({\cal X}_1)G_{r r_2}^{\alpha
\alpha_2}({\cal X}_1,{\cal X}_2)=
-i \hbar \delta({\cal X}_1-{\cal X}_2) 
\delta_{\alpha_1 \alpha_2} \delta_{r_1 r_2}+  \\ &&
+{\Big [}-{{u} \over 2} \delta_{r_1 R} \sum_{\alpha} 
G_2({{\cal X}_1}_R^{\alpha}, {{\cal X}_1}_L^{\alpha_1},
{{\cal X}_2}_{r_2}^{\alpha_2},{{\cal X}_1}_L^{\alpha})
+({{u} \over 4}-u_f) \delta_{r_1 R}  \sum_{\alpha} 
G_2({{\cal X}_1}_R^{\alpha_1},{{\cal X}_1}_L^{\alpha},
{{\cal X}_2}_{r_2}^{\alpha_2},{{\cal X}_1}_L^{\alpha}) {\Big ]}
+(R \leftrightarrow L).
\nonumber
\label{green}
\end{eqnarray}

\vspace{0.1truein}
Because momentum conservation implies
$G_{RL}^{\alpha_1 \alpha_2}=0$, for simplicity we will denote 
$G_{r r}^{\alpha_1 \alpha_2}=G_r^{\alpha_1 \alpha_2}$.

We introduce the self energy for the right- and left- moving particles
by the relation

\vspace{0.1truein}
\begin{equation}
G_{R/L}^{-1 \alpha_1 \alpha_2}({\cal X}_1, {\cal X}_2)=
{G_0}_{R/L}^{-1 \alpha_1 \alpha_2}({\cal X}_1, {\cal X}_2)
-\Sigma_{R/L}^{\alpha_1 \alpha_2}({\cal X}_1, {\cal X}_2),
\label{selfenergydef}
\end{equation}
where $G_0$ is the Green's function in the absence of 
the backscattering interactions between particles.

The self energy can be computed perturbatively. We generalize the method of 
\cite{kadanoff} to include chirality and spin indices. 
In the collision approximation (second order perturbation theory in the
interaction), the self energy is

\vspace{0.1truein}
\begin{eqnarray}
\Sigma_{R/L}^{\alpha_1 \alpha_2}({\cal X}_1, {\cal X}_2)=&&
i {{u} \over {2 \hbar}} G_{L/R}^{\alpha_1 \alpha_2}({\cal X}_1,{\cal X}_2)
\delta({\cal X}_1-{\cal X}_2)
-{i \over {\hbar}} ({{u} \over 4}
-u_f) \sum_{\alpha} G_{L/R}^{\alpha \alpha} ({\cal X}_1,{\cal X}_2)
\delta({\cal X}_1- {\cal X}_2) \delta_{\alpha_1 \alpha_2}
\\&&
+{{u^2} \over {4 \hbar^2}} 
\sum_{\alpha, \beta} G_{L/R}^{\alpha_1 \alpha_2} ({\cal X}_1,{\cal X}_2)
 G_{L/R}^{\alpha \beta} ({\cal X}_2,{\cal X}_1)
G_{R/L}^{\beta \alpha} ({\cal X}_1,{\cal X}_2)
\nonumber \\&&
-{{u (u -4 u_f)} \over {8 \hbar^2}}
\sum_{\alpha, \beta} G_{L/R}^{\alpha_1 \alpha} ({\cal X}_1,{\cal X}_2)
 G_{L/R}^{\alpha \beta} ({\cal X}_2,{\cal X}_1)
G_{R/L}^{\beta \alpha_2} ({\cal X}_1,{\cal X}_2)
\nonumber \\&&
-{{u (u -4 u_f)} \over {8 \hbar^2}} \sum_{\alpha, \beta} 
G_{R/L}^{\alpha_1 \alpha} ({\cal X}_1,{\cal X}_2)
 G_{L/R}^{\alpha \beta} ({\cal X}_2,{\cal X}_1)
G_{L/R}^{\beta \alpha_2} ({\cal X}_1,{\cal X}_2)
\nonumber \\&&
+{{(u-4 u_f)^2} \over {16 \hbar^2}}
 \sum_{\alpha, \beta} G_{R/L}^{\alpha_1 \alpha_2} ({\cal X}_1,{\cal X}_2)
 G_{L/R}^{\alpha \beta} ({\cal X}_2,{\cal X}_1)
G_{L/R}^{\beta \alpha} ({\cal X}_1,{\cal X}_2).
\nonumber
\end{eqnarray}

\subsection{Equations of motion for real time Green's functions}

Following \cite{kadanoff} we perform the analytic continuation
from the imaginary time to real time. We define real time
Green's functions 

\begin{eqnarray}
 g^{\alpha_1 \alpha_2}_r ({\cal X}_1,{\cal X}_2)=
-i < T[ \psi_{r \alpha_1}({\cal X}_1){\psi_{r \beta}^{\dagger}}({\cal X}_2)]>,
\end{eqnarray}

\begin{eqnarray}
{g^{\alpha_1 \alpha_2}_r}^{>}
({\cal X}_1,{\cal X}_2)=-i < \psi_{r \alpha_1}({\cal X}_1)
{\psi_{r \alpha_2}^{\dagger}}({\cal X}_2)>,
\end{eqnarray}

\begin{eqnarray}
{g^{\alpha_1 \alpha_2}_r}^{<}
({\cal X}_1,{\cal X}_2)=i <{\psi_{r \alpha_2}^{\dagger}}({\cal X}_2)
 \psi_{r \alpha_1}({\cal X}_1)>.
\end{eqnarray}

We also define

\begin{equation}
{g_{r}^{\alpha \beta}}(x,t) \equiv
{g_{r}^{\alpha \beta}}(x,t;x,t),
\end{equation}
and we introduce the Fourier transforms:
\begin{eqnarray}
{g_{r}^{\alpha \beta}}^{</>}(p,\omega; x,t) 
\equiv \int dx' \int dt' e^{-i p x'}
e^{i \omega t'} 
 \Big[ \mp i g_{r}^{\alpha \beta}(x+{x' \over 2}, t+{t' \over 2};
x-{x' \over 2}, t-{t' \over 2}) \Big].
\end{eqnarray}

We work in the collision approximation. We also assume that the space and
time dependence of the Green functions of the system 
$g({\cal X}_1,{\cal X}_2)$
vary slowly with respect to the center of mass coordinates $x=(x_1+x_2)/2$
and $t=(t_1+t_2)/2$ \cite{kadanoff}. Thus 
the equations of motion for the real time Green's functions become:

\begin{eqnarray}
\hbar [\partial_t \pm v_F \partial_x]&&
{g_{R/L}^{\alpha_1 \alpha_2}}^{<}({\cal X})=\pm \sum_{\gamma} \Bigg\{
{{u} \over 2}\Big[{g_R^{\alpha_1 \gamma}}^{<}({\cal X}) 
{g_L^{\gamma \alpha_2}}^{<}({\cal X})
-{g_L^{\alpha_1 \gamma}}^{<}({\cal X}) 
{g_R^{\gamma \alpha_2}}^{<}({\cal X})\Big] 
\Bigg\} 
\nonumber \\&&
\pm i {{u^2} \over {8 \hbar}} \sum_{\gamma, \alpha_3,
\alpha_4} \int_{p_i} \int_{\omega_i}  
\bigg\{\Big[{g_L^{\alpha_1 \gamma}}^{<}(p_1, w_1;{\cal X})  
{g_R^{\gamma \alpha_2}}^{>}(p_2, w_2;{\cal X} )
{g_R^{\alpha_4 \alpha_3}}^{<}(p_3, w_3;{\cal X})
 {g_L^{\alpha_3 \alpha_4}}^{>}(p_4, w_4;{\cal X}) \nonumber \\&&
 -(R \leftrightarrow L)\Big] -(< \leftrightarrow >) \bigg\}
2 \pi \delta(p_1+p_3-p_2-p_4) 2 \pi 
\delta(\omega_1+\omega_3-\omega_2-\omega_4)
\nonumber \\&& \mp  
i {{u (u-4 u_f)} \over {16 \hbar}}  \sum_{\gamma, \alpha_3, \alpha_4}
\int_{p_i} \int_{\omega_i}  
\bigg\{\Big[  {g_L^{\alpha_1 \alpha_3}}^{<}(p_1, w_1;{\cal X} ) 
{g_L^{\alpha_3 \alpha_4}}^{>}(p_2, w_2;{\cal X})
{g_R^{\alpha_4 \gamma}}^{<}(p_3, w_3; {\cal X})
 {g_R^{\gamma \alpha_2}}^{>}(p_4, w_4;{\cal X})
\nonumber \\&&
-(R \leftrightarrow L)\Big]
-(< \leftrightarrow >) \bigg\}
2 \pi \delta(p_1+p_3-p_2-p_4) 
2 \pi \delta(\omega_1+\omega_3-\omega_2-\omega_4). 
\label{eqforg}
\end{eqnarray}

Here we denoted $\int_{-\infty}^{\infty} {dp \over {2 \pi}}$ by 
$\int_p$, and similarly for $\omega$. As before we chose
${\cal X}$ to denote $(x,t)$. For simplicity we set 
the high energy cutoff in the momentum and frequency integrals 
at infinity, as the energy scale
beyond which the Luttinger Liquid physics ceases to be valid
is much larger than the other energy scales involved in the problem.

\subsection{The Quantum Boltzmann Equation}

We introduce the generalized Wigner distribution functions by:

\begin{eqnarray}
f^{\alpha_1 \alpha_2}_{r}(p;x,t)=&&\int dx' e^{-i p x'}
[-i{g^{\alpha_1 \alpha_2}_{r}}^{<}(x+{x' \over 2},t;x-{x' \over 2}, t)]
=\int_{\omega}
[{g^{\alpha_1 \alpha_2}_{r}}^{<}(p,\omega; x, t)].
\end{eqnarray}

The equations for these functions can be 
derived from Eq.(\ref{eqforg}) realizing that \cite{kadanoff}:

\begin{eqnarray}
{g_r^{\alpha \beta}}^{<}(p,\omega;{\cal X})=f_r^{\alpha \beta}(p;{\cal X}) 
a_r^{\alpha \beta}(p, \omega),
\end{eqnarray}

\begin{eqnarray}
{g_r^{\alpha \beta}}^{>}(p,\omega;{\cal X})=[\delta_{\alpha \beta}
-f_r^{\alpha \beta}(p;{\cal X})] 
a_r^{\alpha \beta}(p, \omega),
\end{eqnarray}

\begin{eqnarray}
a_{L/R}^{\alpha \beta}(p, \omega)=2 \pi \delta(\omega \pm v_F p).
\end{eqnarray}

We thus obtain:

\begin{eqnarray}
(\partial_t \pm v_F \partial_x)\int_p f_{R/L}^{\alpha_1 \alpha_2}(p)
&&=\pm {{u} \over {2 \hbar}}
i \sum_{\gamma} \int_{p_1,p_2} [f_R^{\alpha_1 \gamma}(p_1) 
f_L^{\gamma \alpha_2}(p_2)
-f_L^{\alpha_1 \gamma}(p_1) f_R^{\gamma \alpha_2}(p_2)] 
\nonumber \\&&
\pm  {{u^2} \over {8 \hbar^2 v_F}} \sum_{\gamma, \alpha3, \alpha_4}
\int_{p_1, p_2} \bigg\{ 
\bigg[f_L^{\alpha_1 \gamma} (p_1) 
[\delta_{\gamma \alpha_2}- f_R^{\gamma \alpha_2}(p_2)] 
f_R^{\alpha_4 \alpha_3}(p_2) 
[\delta_{\alpha_3 \alpha_4} -f_L^{\alpha_3 \alpha_4}(p_1)]
\nonumber \\&&
-[\delta_{\alpha_1 \gamma}- f_L^{\alpha_1 \gamma} (p_1)]
f_R^{\gamma \alpha_2}(p_2 )
[\delta_{\alpha_4 \alpha_3} -f_R^{\alpha_4 \alpha_3}(p_2)]
f_L^{\alpha_3 \alpha_4}(p_1)  \bigg]
-(R \leftrightarrow L) \bigg\}.
\label{eqforf}
\end{eqnarray}

Using the same approximation of slowly varying distributions functions
we can write these equations so they describe
the change in the distribution functions at each momentum 
$p$:

\vspace{0.2truein}

\begin{eqnarray}
(\partial_t + v_F \partial_x)&&f_{R}^{\alpha_1 \alpha_2}(p)
= {{u} \over {2 \hbar}}
i [f_R^{\alpha_1 \gamma}(p) 
\int_{q} f_L^{\gamma \alpha_2}(q)
-\int_{q} f_L^{\alpha_1 \gamma}(q) f_R^{\gamma \alpha_2}(p)] 
\label{eqp1}\\&&
+{{u^2} \over {8 \hbar^2 v_F}} 
\bigg\{ 2 f_R^{\alpha_1 \alpha_2}(p)  \bigg[ f_R^{\alpha_4 
\alpha_3}(p) 
\int_{q} f_L^{\alpha_3 \alpha_4}(q)-\int_q f_L^{\gamma \gamma}
(q)
 \bigg]
-2\int_q f_L^{\alpha_1 \alpha_2}(q)  
\bigg[ f_L^{\alpha_4 \alpha_3}(q) 
 f_R^{\alpha_3 \alpha_4}(p)-f_R^{\gamma \gamma}(p) \bigg]
\nonumber \\&&
+\int_q f_L^{\alpha_1 \gamma}(q) f_R^{\gamma \alpha_2}(p)
\bigg[f_L^{\alpha_3 \alpha_3}(q;{\cal X})-f_R^{\alpha_3
 \alpha_3}(p)\bigg]
-\int_q f_R^{\alpha_1 \gamma}(p) f_L^{\gamma \alpha_2}(q)
\bigg[f_R^{\alpha_3 \alpha_3}(p)-f_L^{\alpha_3 
\alpha_3}(q)\bigg]
\bigg\},
\nonumber
\end{eqnarray}

\begin{eqnarray}
(\partial_t - v_F \partial_x)&&f_{L}^{\alpha_1 \alpha_2}(p)
={{u} \over {2 \hbar}}
i [f_L^{\alpha_1 \gamma}(p) 
\int_{q} f_R^{\gamma \alpha_2}(q)
-\int_{q} f_R^{\alpha_1 \gamma}(q) f_L^{\gamma \alpha_2}(p)] 
\label{eqp2}\\&&
+  {{u^2} \over {8 \hbar^2 v_F}} 
\bigg\{ 2 f_L^{\alpha_1 \alpha_2}(p)  
\bigg[ f_L^{\alpha_4 \alpha_3}(p) 
\int_{q} f_R^{\alpha_3 \alpha_4}(q)-
\int_q f_R^{\gamma \gamma}(q) \bigg]
-2\int_q f_R^{\alpha_1 \alpha_2}(q)  
\bigg[ f_R^{\alpha_4 \alpha_3}(q) 
 f_L^{\alpha_3 \alpha_4}(p)-f_L^{\gamma \gamma}(p) \bigg]
\nonumber \\&&
+\int_q f_R^{\alpha_1 \gamma}(q) f_L^{\gamma \alpha_2}(p)
\bigg[f_R^{\alpha_3 \alpha_3}(q)-
f_L^{\alpha_3 \alpha_3}(p)\bigg]
-\int_q f_L^{\alpha_1 \gamma}(p) f_R^{\gamma \alpha_2}(q)
\bigg[f_L^{\alpha_3 \alpha_3}(p)-
f_R^{\alpha_3 \alpha_3}(q)\bigg]
\bigg\}.
\nonumber
\end{eqnarray}

These are the generalized Boltzmann equations for a one dimensional 
spinfull electron gas.

\end{document}